\documentclass[english,pre,twocolumn,superscriptaddress,letter]{revtex4}

\usepackage{amsmath}
\usepackage{amssymb}
\usepackage{ae,aecompl}
\usepackage{babel}

\providecommand{\onlinecite}{\cite}

\begin{document}

\preprint{PRE Manuscript}

\title{Comment on {}``Jamming at zero temperature and zero applied stress:
The epitome of disorder''}

\author{Aleksandar Donev}

\affiliation{\emph{Program in Applied and Computational Mathematics}, \emph{Princeton
University}, Princeton NJ 08544}

\affiliation{\emph{Princeton Materials Institute}, \emph{Princeton University},
Princeton NJ 08544}

\author{Salvatore Torquato}

\email{torquato@electron.princeton.edu}

\affiliation{\emph{Program in Applied and Computational Mathematics},
\emph{Princeton
University}, Princeton NJ 08544}

\affiliation{\emph{Princeton Materials Institute}, \emph{Princeton University},
Princeton NJ 08544}

\affiliation{\emph{Department of Chemistry}, \emph{Princeton University}, Princeton
NJ 08544}

\author{Frank H. Stillinger}

\affiliation{\emph{Department of Chemistry}, \emph{Princeton University}, Princeton
NJ 08544}

\author{Robert Connelly}

\affiliation{\emph{Department of Mathematics}, \emph{Cornell University}, Ithaca
NY 14853}

\begin{abstract}
O'Hern, Silbert, Liu and Nagel [Phys. Rev. E. 68 , 011306 (2003)] (OSLN) 
claim that a special point  $J$  of a "jamming phase diagram" (in density, 
temperature, stress space) is related to random close packing of hard spheres, and 
that it represents, for their suggested 
definitions of jammed and random, the recently introduced maximally random jammed state.
We point out several difficulties with their definitions and question some of their
claims. Furthermore, we discuss the
connections between their algorithm and other hard-sphere 
packing algorithms in the literature.

\end{abstract}
\maketitle

Jammed random packings of hard particles have been 
and continue to be a subject of intense interest. 
The lack of precise definitions of both ``jammed" and ``random" 
have been a hindrance in the field, and recently efforts have emerged that
have attempted to  correct these deficiencies \cite{Torquato_MRJ, Torquato_jammed, Jamming_LP}. 
In particular, employing these new definitions, it has been shown that
the venerable  random close packed (RCP) state is ill-defined  but can be replaced
by the well-defined notion of a maximally random jammed (MRJ) state \cite{Torquato_MRJ}. 
O'Hern, Silbert, Liu and Nagel (OSLN) \cite{Jamming_ZeroT}  have recently written an
interesting paper \cite{Jamming_ZeroT} that proposes a unified view of jamming for a 
variety of physical systems, including hard-sphere packings.

OSLN acknowledge the weaknesses of the 
conventional RCP state as stated
in Ref. \cite{Torquato_MRJ}, but redefine MRJ (while still calling it RCP) by proposing 
new definitions for what constitutes a jammed and random system \cite{footnote_1}.
Given the subtlety of the problem, new definitions
of ``jammed" and ``random" must be held to high
mathematical standards in order to supplant existing ones.
In this Comment, we question whether OSLN's ``cleaner"
definitions for these terms meet such standards. We also
take issue with their claim to have generated
unbiased and universal results of relevance to
random sphere packings. Finally, we discuss the relationships
between their algorithm and  other packing algorithms.

\section{\label{SectionJammed}What is {}``jammed''?}

OSLN question whether the hard-sphere system is ``physical"
and therefore resort to studying particle systems
with soft-sphere interactions to mimic hard-particle packings.
The latter is inherently a geometrical problem.
In fact, there is a simple and rigorous geometrical approach
to jamming in hard-sphere systems that is not only well-defined, but, as we show
below, is closely related to OSLN's jamming point $J$.

Although the hard-sphere potential is an idealization, it is no less
physical than any soft-sphere potential, especially in regards to jamming.
Indeed, the singular nature of the hard-sphere potential is crucial because it
enables one to be precise about the concept of ``jamming.''
Recently, three hierarchically ordered jamming
categories have been introduced \cite{Torquato_jammed}: \emph{local}, 
\emph{collective} and \emph{strict} \emph{jamming}.
Each successive category progressively relaxes the boundary conditions
imposed on the particle displacements. These definitions are very
intuitive and completely \emph{geometric}, and are closely
linked to definitions of {}``rigid'' or {}``stable'' packings appearing
in the mathematics literature \cite{Connelly_disks,Connelly_packings}.

OSLN's definition of jamming simply states that
the configuration of particles is at a stable (strict) energy minimum.
Such a definition is dependent on the particular interparticle potential,
and thus it obscures the relevant packing geometry (exclusion-volume effects). 
Furthermore, the distinctions between
different jamming categories is critical, especially if one is trying
to determine the density of the MRJ state \cite{Jamming_LP_results}.
Specifically, this density will generally be higher the more demanding is the
jamming category. Clearly, OSLN do not distinguish between different degrees
or levels of jamming.  We have recently demonstrated that 
the distinction between collective and strict jamming is important even for
very large packings, especially in two dimensions \cite{Jamming_LP_results} .

For OSLN, a jammed configuration is one where there are no
zero-frequency modes of the Hessian matrix of the total potential energy
with respect to the positions of the particles (the dynamical
matrix), while keeping the periodic unit cell \emph{fixed}. Our definition of
strict jamming relaxes this requirement and includes the lattice vectors
as additional degrees of freedom \cite{Jamming_LP}. As explained
in detail in Ref. \cite{Connelly_Tensegrities}, the Hessian consists
of two parts, a negative definite \emph{stress matrix} and a positive
semidefinite \emph{stiffness matrix}. OSLN's 
definition of jammed means simply that the Hessian
is positive definite at the energy minimum. A precise phrase for this
is \emph{a stable} or \emph{strict (local) energy minimum}, and we
see no point in redefining this elementary concept. In fact,
according to OSLN, any stable energy minimum represents a jammed configuration,
and it is not possible to relate this idea to \emph{packing} concepts
without numerous additional assumptions about the form of the pair potential
and the interparticle distances at the energy minimum.

Although OSLN point out themselves that our definitions
of jamming and MRJ are for \emph{hard-sphere packings}, they claim
to replace them with a {}``cleaner definition,'' which applies \emph{only}
to systems of \emph{soft spheres}. The two definitions
cannot directly be compared as they apply to different systems.  However,
OSLN themselves clearly imply that their {}``jammed'' soft-sphere
systems and {}``jammed'' hard-sphere packings are related, by referring
to other works on hard-sphere systems. For example, they claim a direct
relation between their special point $J$ and RCP of hard spheres
in Section IID of Ref. \cite{Jamming_ZeroT}. 
The basic idea, as OSLN explain, is that one {}``can approach the
hard sphere by making the potential harder and harder...{[}to{]} produce
a limiting hard sphere value''. However, they question whether the
hard-sphere limit is well-defined and {}``would argue that hard spheres
are a singular limit and thus unphysical'' and that 
{}``One should therefore concentrate
on softer potentials for which unambiguous definitions can be constructed.''

To demonstrate that the limit is well-defined, let us first define
a \emph{collectively jammed sphere packing} to be \emph{any nonoverlapping
configuration of hard spheres in which no subset of particles can
continuously be displaced so that its members move out of contact
with one another and with the remainder set} 
(while maintaining nonoverlap) \cite{Torquato_jammed}. 
The following theorem \cite{Connelly_Tensegrities} shows
that near the {}``jamming threshold'' $\phi_{c}$, as defined in
Section IIB in Ref. \cite{Jamming_ZeroT}, the jamming of particle
systems as defined by OSLN is directly related to this definition of
collective jamming in hard-sphere packings:

\noindent\textbf{Theorem:} {\sl Consider an interparticle potential that
is continuous and strictly monotonically decreasing
around $r_{ij}=D$, and vanishes for $r_{ij}>D+\delta$. If in a finite
configuration of particles interacting with such a potential, all
interacting (i.e., closer then $D+\delta$) particles are a distance
$D$ apart, and the configuration is a stable local energy minimum,
then the configuration corresponds to a collectively jammed packing
of hard spheres with diameter $D$.}

If one relaxes the condition that all interacting particles are exactly
at distance $D$ apart and instead asks only that the minimum interparticle
distance be $D$, then for a sufficiently small $\delta$ one can
prove \cite{Connelly_Energy} that the above sphere packing is almost
collectively jammed (i.e., it is trapped in a small neighborhood of
the initial configuration \cite{Jamming_LP}). This theorem implies
that the packings studied in Ref. \cite{Jamming_ZeroT} that are very
slightly above the {}``jamming threshold'' $\phi_{c}$ are indeed
closely related to collectively jammed ideal packings of spheres of
diameter $D=\sigma$ (polydispersity is trivial to incorporate).
All of these considerations call into question the value of a 
definition of jamming that hinges on eigenvalues of dynamical matrices.

Finally, it is important to note that despite the fact that our definition of
collective jamming above calls for virtual displacing (groups of) particles, one can
in fact rigorously test for our hard-particle 
jamming categories using linear programming
\cite{Jamming_LP}, without what OSLN call {}``shifting particles,''
even for very large disordered packings \cite{Jamming_LP_results}.
We have in fact communicated to OSLN the results \cite{footnote_2}
of our algorithm applied to several sample packings provided by them. In short, our
algorithm verified that  OSLN's systems near $\phi_{c}$ were indeed
nearly collectively jammed (within a very small tolerance) when viewed as
packings. However, they were not strictly jammed because OSLN keep
the lattice vectors fixed during energy minimization.

\section{\label{SectionRandom}What is {}``random''?}

We agree that the maximum of an appropriate ``entropic" metric
would be a potentially useful way to characterize the randomness of a packing
and therefore the MRJ state \cite{Anu_order_metrics}.
However, as pointed out in Ref. \cite{Anu_order_metrics},
a substantial hurdle to be overcome
is the necessity to generate all possible jammed states,
or at least a representative sample of such states,  
in an unbiased fashion using a ``universal'' 
protocol in the large-system limit. 
Even if such a protocol could be developed, however, the issue of what weights to assign
the resulting configurations remains.
Moreover, there are other fundamental problems with entropic
measures, as we discuss below, including its significance 
for two-dimensional monodisperse hard disk packings.

According to OSLN, maximally random is defined by {}``where the entropy
of initial states is maximum'' and imply that this is a universal
measure of disorder. It is not clear exactly what the
authors mean by entropy and how (or whether) it can be measured for
{}``initial states''. It is not obvious that one
can  relate the {}``randomness'' of the \emph{final} configurations
(which is what OSLN are analyzing) to that of the \emph{initial} configurations.
It appears OSLN's rationale is that their algorithm goes to the {}``nearest'' energy
minimum from a given initial configuration.
Does this process preserve {}``entropy'' or randomness? Clearly
if one used, for example, global energy minimization, one would obtain
very different results. Furthermore, entropy is a concept inherently
related to \emph{distributions} of configurations. However, one classifies
\emph{particular} final configurations (packings) as random or disordered,
and by considering a given configuration, one can devise a procedure
for quantitatively measuring (using order metrics) how disordered
or ordered it is. This distinction between distributions of configurations
and particular configurations is an important one that OSLN do not
make.

The MRJ state is defined in \cite{Torquato_MRJ} as the jammed state
which minimizes a given order metric $\psi$. OSLN suggest their interpretation
of maximally random as superior because using order metrics {}``will
always be subject to uncertainty since one never knows if one has
calculated the proper order parameter.'' Therefore, OSLN believe that
they have identified the proper, unique, measure of order (related
to entropy). We wish to stress the difference between \emph{well-defined}
and \emph{unique}, as the two seem to be blurred in Ref. \cite{Jamming_ZeroT}.
The MRJ state is well-defined in that for a particular
choice of jamming category and order metric it can be identified unambiguously.
For a finite system, it will consist of a discrete set (possibly one)
of configurations,  becoming more densely populated as the system becomes larger. 
At least for collective and strict
jamming in three dimensions, a variety of sensible order metrics seem
to produce an MRJ state near $\phi\approx0.64$ \cite{Anu_order_metrics}, the
traditionally accepted density of the RCP state.

However, the \emph{density} of the MRJ state should not be confused
with the MRJ \emph{state} itself. 
It is possible to have a rather
ordered packing at this very same density; for example, 
a jammed but diluted  vacancy FCC lattice packing  \cite{Anu_order_metrics}.
This is why the two-parameter description of packings in terms of the density
$\phi$ and order metric $\psi$, as in Ref. \cite{Torquato_MRJ},
is not only useful, but actually necessary.

OSLN's description of order implies a direct
relation between probability densities and randomness, i.e., that
the \emph{most probable} \cite{footnote_3}
configurations represent the most disordered state. In this sense,
one expects that the density of jammed configurations, when viewed
as a three-dimensional plot over the $\phi-\psi$ plane will be very
strongly peaked around the MRJ point for very large systems, just
as the probability distribution curves in Fig. 6 in Ref. \cite{Jamming_ZeroT}
are very peaked around $\phi\approx0.64$. As OSLN suggest themselves,
this might explain why several different packing procedures yield
similar hard-particle packings under appropriate conditions, 
historically designated as RCP.

However, this is far from being a closed question \cite{footnote_4}. 
Consider two-dimensional monodisperse circular disk packings as an example.
It is well-known that two-dimensional analogs of
three-dimensional computational and experimental protocols
that lead to putative RCP states result in disk
packings that are highly crystalline,
forming rather large triangular domains (grains) \cite{footnote_4a}.
 Because such highly ordered packings are the most probable for these
\emph{protocols}, OSLN's entropic measure would identify these as the most disordered, 
a dubious proposition. An appropriate order metric, on the other hand,
is capable of identifying a particular configuration (not an ensemble
of configurations) of considerably lower density
(e.g., a jammed diluted triangular lattice)
that is consistent with our intuitive notions
of maximal disorder. However, typical packing protocols would
almost never generate such disordered disk configurations because
of their inherent bias toward undiluted crystallization.
This brings us to OSLN's claim that they have 
devised an unbiased universal protocol, to which we now turn
our attention.

\section{\label{SectionAlgorithms}Universal, Hard and Soft Algorithms}

In this section, we focus on the algorithms used by OSLN and 
point out why they are neither universal nor superior to other procedures.
We point out the close relations between OSLN's algorithm
for generating configurations near the onset of jamming and
the Zinchenko hard-sphere packing algorithm \cite{Zinchenko}. 
Furthermore, we question OSLN's implication that using
one kind of interaction potential (with three different exponents) 
and one algorithm amounts
to exploring the space of all jammed configurations in an unbiased
manner. This puts into doubt the claimed universality of the point $J$.

By fixing the interaction potential, initial density and energy
minimization (conjugate gradient) algorithm, OSLN obtain a well-defined collection
of final configurations with well-defined (not unique) properties. 
In essence, OSLN make their algorithm devoid of tunable parameters
by simply choosing specific and fixed values for them. 
Both the Zinchenko and OSLN algorithms are {}``dynamics independent'', in the sense
that there is no tunable parameter for the rate of compression, which
would be an analog of the cooling or quenching rate in molecular systems.
Both also imply that this makes their algorithm
universal or superior to other algorithms and that the (well-defined)
results they obtain are somehow special.
Most sensible algorithms will in fact produce
a well-defined density in the limit of large systems given a choice
of algorithmic parameters. For example, by changing the expansion
rate in the Lubachevsky-Stillinger algorithm, one can achieve final
densities for spheres anywhere in the range from $0.64$ (fast expansion) to $0.74$
(very slow expansion), as clearly illustrated in Fig. 2a in Ref. \cite{Torquato_MRJ}.
Therefore, if we followed the logic of OSLN, we could claim that any
number in that range represents a special point. 
In our opinion, a good packing algorithm should be capable of generating
a variety of packings, in both density and the amount of order. How
can one ascertain that the packings one produces are {}``most random''
if there are no other jammed packings to compare to?

OSLN use two main procedures to generate final configurations. 
The first procedure is to choose a density
and then use conjugate gradients to find a nearby energy minimum,
starting from a randomly-generated initial configuration ($T=\infty$),
as described in Section IIA in Ref. \cite{Jamming_ZeroT}.
Using this procedure, OSLN sampled inherent structures \cite{Inherent_Structures}
at fixed density to measure the fraction $f_j(\phi)$
of states that had nonzero bulk and shear moduli, and showed that
$f_j$ has a strong system-size dependence
with its derivative becoming a delta-function in the large system limit.
It is important to note that this procedure as such has little or nothing to do with
hard-sphere packings, especially for the kind of soft potentials ($\alpha\geq3/2)$
that OSLN study. Many stable energy  minima will be completely unrelated
to packings, and especially not to those designated as MRJ states.

OSLN used a second procedure to study the mechanical and structural
properties of systems near the onset of jamming $\phi_{c}$.
In this procedure, a configuration is compressed (or decompressed) 
using very small steps in density until the bulk and shear
moduli vanished (or nonzero moduli develop), as described in Section IIB in Ref. \cite{Jamming_ZeroT}.
We now demonstrate that this procedure 
is closely related to Zinchenko's  algorithm \cite{Zinchenko} 
for generating hard-sphere packings. Start at low density with a set of nonoverlapping spheres
of diameter $\sigma$. Both algorithms then slowly grow the particles (OSLN
in small increments, Zinchenko continuously) while moving the particles
to avoid overlap \cite{footnote_7}. 
In the Zinchenko algorithm, one strictly maintains the contact between
particles as soon as they touch, which requires solving a system of
ODE's containing the rigidity matrix of the packing \cite{Jamming_LP}
to find the necessary particle displacements. OSLN on the other hand,
use conjugate gradients (CG) to reminimize the potential energy, which
will simply push the particles just enough to keep them nonoverlapping,
i.e., almost in contact. This procedure continues in both algorithms
until no further densification is possible without inducing overlap.

Accordingly, it is not surprising
the packing configurations close to $\phi_{c}$
obtained in Ref. \cite{Jamming_ZeroT} closely resemble (in packing
fraction, amorphous character, coordination, etc.) packings generated
via a variety of bonafide  hard-sphere algorithms (and experiments
\cite{Experiment_RCP}). In particular, very similar packings are produced
with the Lubachevsky-Stillinger (LS) algorithm \cite{LS_algorithm,LS_algorithm_3D}
(with sufficiently high expansion rates) and the Zinchenko algorithm
\cite{Zinchenko}. OSLN criticize the LS algorithm 
for changing the density in a dynamic fashion. The stated advantage
of the OSLN protocol is that one can {}``quench the system to the final
state within a fixed energy landscape'' since {}``the density is
always held constant''. We are very puzzled by this last claim in
light of their admission (in Section IIB of Ref. \cite{Jamming_ZeroT})
that they slowly change the density of the packing
to find $\phi_{c}$. In fact, OSLN do not seem to clearly distinguish
between the two rather different procedures they employ: the first
for finding inherent structures (at a fixed density) and  the second
for generating packings at the jamming threshold (which searches in
density). Fig. 6 of Ref. \cite{Jamming_ZeroT}, which
supposedly represents the distributions of jamming thresholds $\phi_{c}$,
\emph{defined} by the \emph{second} procedure, is obtained by differentiating the distribution
generated with the \emph{first} procedure, with no clear justification.

Most problematic of all is OSLN's claim that their results are universal.
Despite the statement that {}``Starting with randomly generated $T=\infty$
states guarantees that we sample \emph{all} {[}emphasis added{]} phase
space equally'', all that their first algorithm manages to explore
is the space of energy minima for the \emph{particular} chosen interaction
potential. By comparing three different exponents
$\alpha$, OSLN conclude that the exact form of the potential is not
important. However, a much more convincing picture would have been
made if they instead tried \emph{qualitatively} different kinds
of interaction potentials, 
rather then simply changing the curvature of the potential at the
contact point. Otherwise, why focus on continuous interaction potentials
at all? Since it is geometry (i.e., the nonoverlap condition on
the spherical cores) that is crucial, the hard-sphere system
offers a far {}``cleaner'' system to study when trying to understand
the special point $J$.


\begin{thebibliography}{24}
\bibitem[{foo({\natexlab{b}})}]{footnote_2}
\bibinfo{note}{Some of these can be found on our webpage \cite{webpage}.}
\bibitem[{foo({\natexlab{d}})}]{footnote_4}
\bibinfo{note}{More fundamental questions have not yet been settled. For
  example, should one really ignore dynamics 
  when trying to account for the well-known density of $0.64$?}
\bibitem[{foo({\natexlab{e}})}]{footnote_4a}
\bibinfo{note}{OSLN's procedure is not immune to this tendency to crystallize,
  which is the reason why they chose to study bidisperse disk packings instead
  of monodisperse disk packings.}
\bibitem[{foo({\natexlab{f}})}]{footnote_7}
\bibinfo{note}{OSLN sometimes use the reverse procedure in which they decrease
  the density, which does not directly relate to the Zinchenko algorithm.}
\end{thebibliography}
\end{document}